\documentclass[11pt]{article}

\usepackage[preprint]{acl}

\usepackage{times}
\usepackage{latexsym}
\usepackage{enumitem}

\usepackage[T1]{fontenc}

\usepackage[utf8]{inputenc}

\usepackage{microtype}

\usepackage{inconsolata}

\usepackage{graphicx}

\usepackage{tikz}
\usetikzlibrary{positioning,arrows.meta}

\usepackage{listings}
\usepackage{xcolor}

\usepackage{booktabs}

\usepackage{placeins}

\usepackage{comment}

\definecolor{diffadded}{RGB}{220,235,255}
\definecolor{diffremoved}{RGB}{255,235,210}
\definecolor{diffaddedtext}{RGB}{0,70,150}
\definecolor{diffremovedtext}{RGB}{200,100,0}

\lstdefinestyle{codestyle}{
    basicstyle=\ttfamily\footnotesize,
    breaklines=true,
    columns=flexible,
    keepspaces=true,
    showstringspaces=false,
    frame=single,
    xleftmargin=0.5em,
    xrightmargin=0.5em,
}

\newcommand{\codeadded}[1]{\colorbox{diffadded}{\textcolor{diffaddedtext}{\ttfamily #1}}}
\newcommand{\coderemoved}[1]{\colorbox{diffremoved}{\textcolor{diffremovedtext}{\ttfamily #1}}}
\title{VICBench: A Multi-Language Benchmark for Code Vulnerability Detection}

\author{
  \textbf{Jin Lu}\textsuperscript{1},
  \textbf{Xuening Han}\textsuperscript{1},
  \textbf{Yang Zhong}\textsuperscript{1,2},
  \textbf{Lin Tan}\textsuperscript{1,3},
\\
  \textbf{Kevin Luo}\textsuperscript{1},
  \textbf{Andrew Gacek}\textsuperscript{1},
  \textbf{Neha Rungta}\textsuperscript{1}
\\
  \\[-0.2em]
   \textsuperscript{1}Amazon Web Services \quad \textsuperscript{2}University of Pittsburgh \quad \textsuperscript{3}Purdue University \\
   \\[-0.5em]
  \texttt{jinlu2@asu.edu} \quad
  \texttt{xuenihan@amazon.com} \quad
  \texttt{yaz118@pitt.edu} \quad
  \texttt{lintan@purdue.edu}
}

\begin{document}
\maketitle
\begin{abstract}
Evaluating security vulnerability detection tools requires benchmark datasets with vulnerability-inducing commits (VICs)---the commits that first introduce vulnerabilities into codebases.  VICs are essential for determining the full range of vulnerable software versions.  
Existing vulnerability datasets suffer from limited programming language coverage, restricted patch complexity, and narrow project scope. 
Through our dual annotation by human experts and an agentic workflow, we create a benchmark---\emph{VICBench}---of 100 verified VICs for 100 CVEs across 88 projects in Python, Java, and C++, covering 48 CWE types. 
VICBench features complex real-world vulnerability fixes averaging 38.6 lines and corresponding VICs of 252.5 lines---significantly larger than prior work. 
Our evaluation shows that state-of-the-art algorithms V-SZZ 
and LLM4SZZ achieve only 33.3\%–40.1\% F1, confirming that using existing approaches still entails significant manual effort. VICBench enables robust evaluation of vulnerability detection approaches.
\end{abstract}

\section{Introduction}

Software vulnerability detection is critical for system security. Proactive detection---identifying vulnerabilities before they reach production---has become increasingly important as detection approaches evolve from static analyzers to deep learning models~\citep{lin2020deepvulsurvey,steenhoek2023empirical,rahman2024causalvul} and LLM-powered code review assistants~\citep{li2025iris,anthropic2025security}. Robust evaluation of these diverse approaches requires high-quality benchmark datasets with vulnerability-inducing commits (VICs). Following~\citet{bao2022vszz}, we define VICs as the commits that first introduce vulnerabilities into codebases. VICs mark the beginning of vulnerability lifecycles and represent the transition from non-vulnerable to exploitable code, enabling evaluation on original vulnerability patterns before refactoring obscures the initial flaw.

A comprehensive benchmark requires repository context, CVE descriptions, fixing commits, and critically, VICs. While the National Vulnerability Database~\citep{nvd2026} provides CVE descriptions and fixes, VICs are not documented. This allows assessing whether tools can detect vulnerabilities at their source and tracking detection capability throughout vulnerability evolution. Without VICs, benchmarks can only evaluate tools on evolved vulnerability forms before fixes, missing evaluation on the earliest and most critical vulnerability patterns.

Existing VIC datasets have significant limitations. V-SZZ~\citep{bao2022vszz} verified 172 CVEs but restricted annotations to 1-5 deleted lines, excluding complex multi-file patches. \citet{jiang2024linuxcve} contributed 1,000+ VICs but focused solely on Linux kernel (C/C++), limiting generalizability. \citet{chen2025vulnerabilityaffectedversionsidentificationfar} annotated 1,128 C/C++ vulnerabilities at high cost (0.5 person-hours each) but did not release VIC annotations. Modern approaches require greater language diversity, patch complexity, and project coverage. Moreover, automated VIC detection faces challenges. Tracing-based methods such as SZZ~\citep{SZZ2005} and variants~\citep{bao2022vszz,tang2025llm4szz} use \texttt{git-blame} heuristics, which struggle with file refactoring, cross-file code movement, and complex vulnerability semantics. Unlike regular bugs, vulnerabilities persist longer due to fundamental design flaws~\citep{nguyen2016vulversion} and often involve addition-only fixes without deleted lines~\citep{chen2025vulnerabilityaffectedversionsidentificationfar}, making traditional approaches ineffective, with studies showing SZZ variants achieve as low as \textbf{9\% precision} and a best-case F-measure of only 61\% even on regular bugs~\citep{rosa2021szzoracle}.
Our contributions include:
\begin{itemize}[leftmargin=12pt, noitemsep]
    \item A benchmark of \textbf{100 verified VICs} for 100 CVEs across 88 projects, 3 programming languages (Python, Java, and C++), and 48 CWE types. Our dataset features significantly larger patches than prior work: fix commits average 38.6 lines and VICs average 252.5 lines.
    \item A dual-annotation approach with human expert verification and agentic workflow assistance achieving $\kappa$=0.707 agreement, ensuring data quality and scalable dataset construction.
    \item An evaluation showing that existing VIC detection algorithms achieve only 33.3\%--40.1\% F1, indicating that existing automated approaches are not substitutes for manual annotation.
\end{itemize}

\noindent\textbf{Availability}: \href{https://zenodo.org/records/18944736?preview=1&token=eyJhbGciOiJIUzUxMiJ9.eyJpZCI6ImMyZmVlOGRhLTM2MzctNGRkNy1iZTgxLTc1NGY1MmMxODc0MCIsImRhdGEiOnt9LCJyYW5kb20iOiI5ZWI0M2Q1OTI0YWVhM2FlMTkxZTIwZDc2OTNhZjA1MyJ9.ahn2MeFMLKLHnMgmVvgj-GfT2X-wHbhg8wL9woRANIfYBEwyVUIAUJkQ--6WSQ0xlyJ1c_-4gEemVlZVPLKeRA}{https://zenodo.org/records/18944736}
\section{Background and Related Work}\label{sec:related_work}

\paragraph{Vulnerability-Inducing Commit Detection}
Tracing-based approaches for identifying VICs build on the {Śliwerski}-Zimmermann-Zeller (SZZ) algorithm~\citep{SZZ2005}, originally designed for bug-inducing commits. \textsc{VCCFinder}~\citep{perl2015vccfinder} first adapted SZZ to vulnerabilities, followed by variants including \textsc{AG-SZZ}~\citep{kim2006agszz}, \textsc{MA-SZZ}~\citep{dacosta2017maszz}, and \textsc{RA-SZZ}~\citep{neto2018raszz} evaluated in V-SZZ~\citep{bao2022vszz}. Recent approaches like \textsc{SEM-SZZ}~\citep{tang2024semszz} use semantic analysis with data and control flow, while \textsc{Vercation}~\citep{cheng2025vercation} and \textsc{LLM4SZZ}~\citep{tang2025llm4szz} incorporate LLMs for improved analysis. However, these methods struggle with file refactoring, cross-file code movement, and addition-only fixes~\citep{chen2025vulnerabilityaffectedversionsidentificationfar}. Concurrent work by~\citet{shi2026agenticzz} addresses some of these limitations using graph-based reasoning for general bug-inducing commits, though without released datasets or security-specific focus. These challenges motivate the need for manually curated benchmarks that capture real-world complexity.

\paragraph{Vulnerability-Inducing Datasets}
Given the distinction between bug-inducing and vulnerability-inducing tasks~\citep{camilo2015bugvul}, several VIC datasets have been developed. VCCFinder~\citep{perl2015vccfinder} constructed 718 CVEs but was found unreproducible~\citep{riom2021replicatevccfinder}. V-SZZ~\citep{bao2022vszz} manually verified 172 CVEs across 46 projects but restricted annotations to fixes with 1-5 deleted lines, limiting coverage of complex patches. \citet{jiang2024linuxcve} contributed 1,000+ VICs but focused solely on Linux kernel (C/C++), limiting language diversity. Concurrent work by~\citet{chen2025vulnerabilityaffectedversionsidentificationfar} curated 1,128 C/C++ vulnerabilities from nine projects at high annotation cost (0.5 person-hours each) but did not release VIC annotations. Our dataset addresses these limitations by covering 100 CVEs across three programming languages (Python, Java, and C++) and 88  projects with significantly larger and more complex patches (Table~\ref{tab:dataset_comparison}), providing a robust publicly-available benchmark for evaluating vulnerability detection approaches.

\section{Motivating Example}
\label{sec:motivating_example}

\begin{figure}[h]
\centering
\small
\begin{tikzpicture}[scale=0.65, transform shape,
    node distance=0.5cm,
    every node/.style={font=\small},
    commit/.style={rectangle, draw, minimum width=2.6cm, minimum height=1cm, align=center, fill=white},
    vic/.style={commit, fill=red!20, draw=red!60!black, thick},
    refactor/.style={commit, fill=blue!10, draw=blue!50!black},
    algo/.style={commit, fill=yellow!20, draw=orange!70!black, thick},
    fix/.style={commit, fill=green!20, draw=green!60!black, thick},
    timeline/.style={->, thick, gray!60},
    label/.style={font=\small, text width=2.6cm, align=center}
]

\draw[timeline, line width=1pt] (0,0) -- (10,0);

\foreach \x/\year in {0/2016, 4/2018, 10/2021} {
    \draw[gray!60] (\x,0.12) -- (\x,-0.12);
    \node[below, font=\small] at (\x,-0.3) {\year};
}

\node[vic] (vic) at (0.5, 2.3) {\textbf{c06d1db}\\Aug 2016};
\node[label, below=0.04cm of vic, text=red!70!black] {\textbf{Initial VIC}\\\texttt{OidcProfileCreator}};

\node[algo] (refactor) at (4.5, 3.8) {\textbf{d60a490}\\Nov 2018};
\node[label, below=0.04cm of refactor, text=orange!80!black] {\textbf{V-SZZ \&}\\\textbf{LLM4SZZ}\\\textbf{stopped here}\\to \texttt{TokenValidator}};

\node[fix] (fix) at (9.5, 2.3) {\textbf{22b82ff}\\Dec 2021};
\node[label, below=0.04cm of fix, text=green!70!black] {\textbf{Security Fix}\\Added validation};

\draw[->, thick, red!60!black] (vic) -- (refactor);
\draw[->, thick, orange!70!black, dashed] (refactor) -- (fix);

\draw[very thick, red!50, dashed, line width=1.5pt] (0.5,0.5) -- (9.5,0.5);
\node[fill=white, font=\small, text=red!70!black] at (5, 0.5) {Vulnerability Persists};

\end{tikzpicture}
\caption{CVE-2021-44878 evolution timeline. V-SZZ and LLM4SZZ stopped at the 2018 refactoring commit (d60a490) where code moved to a new file, missing the true VIC (c06d1db) from 2016.}
\label{fig:cve_timeline_44878}
\end{figure}
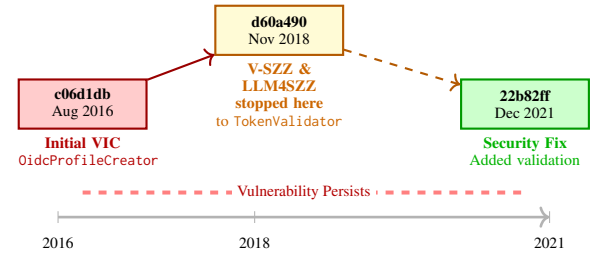

CVE-2021-44878, a pac4j vulnerability accepting unsigned ID tokens, demonstrates cross-file code movement defeats git-blame approaches (Figure~\ref{fig:cve_timeline_44878}). 

\paragraph{Step 1: Initial Introduction (c06d1db, Aug 2016)} The vulnerability was introduced in \texttt{OidcProfileCreator.java}, accepting unsigned tokens without validation:

\begin{lstlisting}[style=codestyle, basicstyle=\ttfamily\scriptsize]
if ("none".equals(jwsAlgorithm.getName())) {
    jwsAlgorithm = null;  // UNSAFE: Accept unsigned tokens
} ...
if (jwsAlgorithm == null) {
    this.idTokenValidator = new IDTokenValidator(
        issuer, clientID);  // No signature verification!}
\end{lstlisting}

\paragraph{Step 2: Code Movement (d60a490, Nov 2018)} Code extracted to \texttt{TokenValidator.java}, preserving the vulnerable pattern:

\begin{lstlisting}[style=codestyle, basicstyle=\ttfamily\scriptsize]
if ("none".equals(jwsAlgorithm.getName())) {
    jwsAlgorithm = null;  // STILL UNSAFE
} ...
if (jwsAlgorithm == null) {
    idTokenValidator = new IDTokenValidator(
        issuer, clientID);  // STILL no verification!}
\end{lstlisting}

\paragraph{Step 3: Security Fix (22b82ffd, Dec 2021)} Added validation to reject unsigned tokens:

\begin{lstlisting}[style=codestyle, basicstyle=\ttfamily\scriptsize, escapechar=@]
@\coderemoved{-if ("none".equals(jwsAlgorithm.getName())) \{}@
@\coderemoved{-~~~~jwsAlgorithm = null;}@
@\coderemoved{-\}}@

final IDTokenValidator idTokenValidator;
@\coderemoved{-if (jwsAlgorithm == null) \{}@
@\codeadded{+if ("none".equals(jwsAlgorithm.getName())) \{}@
@\codeadded{+~~~~if (!configuration.isAllowUnsignedIdTokens()) \{}@
@\codeadded{+~~~~~~~~throw new TechnicalException(...)}@
@\codeadded{+~~~~logger.warn("Allowing unsigned ID tokens");}@
    idTokenValidator = new IDTokenValidator(...);
\end{lstlisting}

\paragraph{Git Blame Limitation} Both V-SZZ and LLM4SZZ rely solely on \texttt{git blame}, which returns commit d60a490 with ``New File'' markers when tracing \texttt{TokenValidator.java}. Git blame cannot track cross-file code movement---only file renames---causing both algorithms to misidentify the refactoring commit as the vulnerability origin.

\paragraph{Pattern-Based Identification} VIC-Agent combines \texttt{git blame} with LLM-guided pattern search using \texttt{git log -S} to cross file boundaries. By extracting the vulnerable pattern and searching across all files and history, it found two candidates: d60a490 (2018) in \texttt{TokenValidator.java} and c06d1db (2016) in \texttt{OidcProfileCreator.java}. It classified d60a490 as refactoring and verified c06d1db as the true VIC.

This example demonstrates that VIC detection requires reasoning about vulnerability semantics and code evolution across file boundaries, motivating the need for manually curated benchmarks. An additional case study (CVE-2019-15477) in Appendix~\ref{sec:appendix_case_1} illustrates complementary challenges.

\section{Benchmark Creation}
\label{sec:data_creation}

\subsection{Annotation Approach}

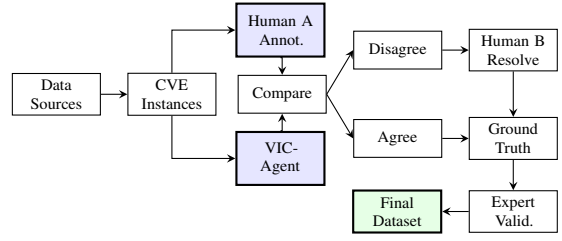
\begin{figure}[h]
\centering
\scriptsize
\begin{tikzpicture}[
    node distance=0.25cm and 0.35cm,
    box/.style={rectangle, draw, text width=1cm, align=center, minimum height=0.5cm, font=\tiny},
    proc/.style={rectangle, draw, thick, text width=1cm, align=center, minimum height=0.7cm, fill=blue!10, font=\tiny},
    result/.style={rectangle, draw, thick, text width=1cm, align=center, minimum height=0.5cm, fill=green!10, font=\tiny},
    arrow/.style={->, >=stealth, thin}
]

\node[box] (sources) at (0,0) {Data\\Sources};
\node[box] (cves) [right=0.35cm of sources] {CVE\\Instances};
\node[proc] (human) [above right=0.2cm and 0.25cm of cves] {Human A\\Annot.};
\node[proc] (agent) [below right=0.2cm and 0.25cm of cves] {VIC-\\Agent};
\node[box] (agree) [right=0.28cm of cves] {Compare};

\node[box] (split2) [above right=0.07cm and 0.35cm of agree] {Disagree};
\node[box] (split1) [below right=0.07cm and 0.35cm of agree] {Agree};

\node[box] (resolve) [right=0.35cm of split2] {Human B\\Resolve};

\node[box] (gt) [right=0.35cm of split1] {Ground\\Truth};

\node[box] (expert) [below=0.4cm of gt] {Expert\\Valid.};
\node[result] (final) [left=0.35cm of expert] {Final\\Dataset};

\draw[arrow] (sources) -- (cves);
\draw[arrow] (cves) |- (human);
\draw[arrow] (cves) |- (agent);
\draw[arrow] (human.south) -| (agree.north);
\draw[arrow] (agent.north) -| (agree.south);
\draw[arrow] (agree.east) -- (split2.west);
\draw[arrow] (agree.east) -- (split1.west);
\draw[arrow] (split2) -- (resolve);
\draw[arrow] (split1) -- (gt);
\draw[arrow] (resolve) -- (gt.north);
\draw[arrow] (gt) -- (expert);
\draw[arrow] (expert) -- (final);

\end{tikzpicture}
\caption{Annotation workflow: data collection, independent annotation procedures (human and agentic), disagreement resolution, and expert validation.}
\label{fig:annotation_workflow}
\end{figure}

To ensure accurate labeling of VICs, 
we design a workflow (Figure~\ref{fig:annotation_workflow}) in which two independent procedures identify VICs for each vulnerability, and their results are then cross-validated to 
resolve discrepancies.

\subsubsection{Procedure 1: Human Annotation}

One author with nine years of programming experience annotated VICs for all 100 instances. Following~\citet{RodriguezPerez2020How} and V-SZZ~\citep{bao2022vszz}, the annotator iteratively traced modified lines using CVE/CWE descriptions, commit messages, and git blame. While V-SZZ identifies the \textit{earliest} commit containing vulnerable lines~\citep{bao2022vszz}, our annotation focuses on when the vulnerability as a \textit{security flaw} was semantically introduced—critical for cases involving refactoring or code movement. This required: (1) examining function names and program logic for addition-only fixes; (2) tracking only security-relevant lines.

\subsubsection{Procedure 2: VIC-Agent}

To validate the robustness of our manual annotations and enable scalable dataset construction, we developed VIC-Agent, an automated workflow powered by large language models. Unlike traditional SZZ algorithms that rely on fixed heuristics, VIC-Agent brings critical flexibility through adaptive decision-making at each step, dynamically selecting between git tools (\texttt{git blame}, \texttt{git log -S}, \texttt{git show}) and employing LLM-based vulnerability analysis and commit verification.

The agent first analyzes the CVE description and fixing commit to understand the vulnerability's \textit{security semantics}. When examining candidate commits, it uses LLM reasoning to verify whether a commit \textit{introduced} new vulnerable logic or merely \textit{refactored} existing vulnerable code. For refactored commits, it continues searching backwards through the pattern evolution chain.

This enables VIC-Agent to handle real-world complexities—file renames, cross-file movements, and refactoring patterns—that static algorithms cannot address. VIC-Agent achieved 84.4\% recall and 94.9\% precision in agreement with human annotations (see Section~\ref{sec:data_analysis}), reflecting its role as a construction tool and validating the dataset quality.

\subsection{Quality Validation}

 Two annotations were considered in agreement if and only if both identified the exact same commit.
 
 \paragraph{Inter-Annotator Agreement:} Human A and VIC-Agent achieved 71\% observed agreement (71 of 100 cases). Computing Cohen's kappa with chance agreement correction (see Appendix~\ref{sec:appendix_kappa} for details), we obtained $\kappa = 0.707$, indicating \textit{substantial agreement}. The observed agreement (71\%) far exceeds chance expectation (1.15\%).

\paragraph{Disagreement Resolution:} For the 29 disagreement cases, a second human annotator (Human B) independently reviewed each instance, examining commit context, vulnerability semantics, and code evolution history. Human B incorporated insights from Human A's annotations and VIC-Agent's reasoning to establish the ground truth annotations.

\paragraph{Expert Validation:} To further validate dataset quality, a principal security engineer with 16 years of security experience independently reviewed a random sample of 11 instances from the final dataset (11\%). This validation confirmed the correctness of the vulnerability-inducing commit identifications and provided additional confidence in the dataset's reliability.

\subsection{Data Sources}

We randomly sampled 100 instances from 88 projects across Python, Java, and C/C++ from three datasets: (1) \textbf{ReposVul}~\citep{wang2024reposvul}, filtered for entries with CVE-relevant files, (2) \textbf{CWE-Bench-Java}~\citep{li2025iris}, and (3) \textbf{VJBench}~\citep{wu2023vjbench}. Our data set cover diverse CWE and patch types.
%
\section{Dataset Analysis}
\label{sec:data_analysis}

\subsection{Dataset Composition and Characteristics}

\paragraph{Vulnerability Type Coverage:} Our dataset encompasses 48 unique CWE types across 100 CVEs, with Cross-site Scripting (CWE-79, 12\%) and Path Traversal (CWE-22, 10\%) being most prevalent (see Appendix~\ref{sec:appendix_cwe_dist}). This diversity ensures comprehensive evaluation across varied security contexts.

\paragraph{Patch Complexity:} Table~\ref{tab:dataset_comparison} compares VICBench with existing benchmarks. Fix commits average 38.6 lines changed (median: 22), while VICs average 252.5 lines (median: 181). This 6.5X difference reflects real-world scenarios where vulnerabilities occur in larger feature implementations.

\begin{table}[h]
\centering
\footnotesize
\setlength{\tabcolsep}{2.5pt}
\begin{tabular}{lccccc}
\toprule
\textbf{Dataset} & \textbf{N} & \textbf{Lang} & \textbf{Prj} & \textbf{Fix/VIC} & \textbf{Pub} \\
\midrule
V-SZZ~\citeyearpar{bao2022vszz} & 172 & C/Ja & 46 & 1-5 / - & Y \\
Jiang et al.~\citeyearpar{jiang2024linuxcve} & 1000+ & C & 1 & - / - & Y \\
Chen et al.~\citeyearpar{chen2025vulnerabilityaffectedversionsidentificationfar} & 1128 & C & 9 & Div / - & N \\
\textbf{Ours} & \textbf{100} & \textbf{Py/Ja/C} & \textbf{88} & \textbf{39 / 253} & \textbf{Y} \\
\bottomrule
\end{tabular}
\caption{Dataset comparison. Fix/VIC Lines show mean changed lines. Pub: Y/N indicates publicly available/unavailable. Our dataset has larger patches, languages (Python/Java/C++), and more projects than prior work.}
\label{tab:dataset_comparison}
\end{table}

\subsection{VIC Detection Performance and Dataset Construction Validation}

To demonstrate that our dataset could not be trivially acquired by running existing algorithms, we evaluated V-SZZ~\citep{bao2022vszz} (Java/C++ subset, 40 CVEs) and LLM4SZZ~\citep{tang2025llm4szz} (all 100 instances). Table~\ref{tab:baseline_performance} shows that V-SZZ achieves 33.3\% F1 and LLM4SZZ achieves 40.1\% F1, demonstrating that our dataset captures complexities beyond current automated approaches. Accurate VIC identification has practical implications for version range determination: \citet{bao2022vszz} found that 99 out of 172 CVEs had spurious version information in NVD and proposed an improved SZZ algorithm to address this. However, cases involving code refactoring and cross-file movement remain challenging—for example, in our motivating example of CVE-2021-44878 (Section~\ref{sec:motivating_example}), identifying the 2018 refactoring rather than the 2016 true VIC would incorrectly exclude two years of vulnerable versions. Our VIC-Agent achieves 89.3\% F1 by focusing on semantic vulnerability introduction, extending prior work to handle such complex scenarios.

Note that VIC-Agent's performance should be interpreted as a construction-tool reference ceiling rather than an independent evaluation, as it participated in the annotation process~\citep{cao2025rigor}.

\begin{table}[h]
\centering
\small
\begin{tabular}{lccc}
\toprule
\textbf{Method} & \textbf{Recall} & \textbf{Precision} & \textbf{F1 Score} \\
\midrule
V-SZZ\textsuperscript{\dag} & 52.3\% & 24.5\% & 33.3\% \\
LLM4SZZ & 56.9\% & 31.0\% & 40.1\% \\
VIC-Agent & 84.4\% & 94.9\% & 89.3\% \\
\bottomrule
\end{tabular}
\caption{VIC detection performance. \textsuperscript{\dag}V-SZZ on Java/C++ only (40 CVEs), LLM4SZZ and VIC-Agent on all 100 instances.}
\label{tab:baseline_performance}
\end{table}

\section{Conclusion}

We present a benchmark of 100 verified vulnerability-inducing commits across Python, Java, and C++, addressing critical gaps in existing VIC datasets. Through dual annotation achieving $\kappa$=0.707 agreement, our dataset captures real-world complexity with significantly larger patches than prior work. Evaluation shows state-of-the-art algorithms achieve only 33.3\%--40.1\% F1, confirming the necessity of manual curation. This benchmark enables robust evaluation of vulnerability detection approaches and supports research in proactive security tooling.

\section*{Limitations}

While our dataset advances VIC benchmarking, several limitations should be acknowledged:

\paragraph{Dataset Scale and Coverage:} Our dataset contains 100 manually verified CVEs, which is smaller than some automatically constructed datasets but prioritizes annotation quality over quantity. The dataset covers three programming languages (Python, Java, C++), which may not generalize to other languages such as JavaScript, Go, or Rust. Additionally, C++ is underrepresented at 8\% (8 CVEs), reflecting its limited availability in our source datasets (ReposVul, CWE-Bench-Java, VJBench) rather than a deliberate design choice; this may limit the benchmark's utility for evaluating tools targeting C/C++ codebases specifically. Future work could expand coverage to additional languages and larger scale while maintaining annotation quality.

\paragraph{Data Source Selection:} Our CVEs were sampled from existing curated datasets (ReposVul, CWE-Bench-Java, VJBench), which may introduce sampling bias toward certain project types, vulnerability patterns, or temporal distributions. The dataset focuses on open-source projects from public repositories, and findings may not fully generalize to proprietary or closed-source codebases.

\paragraph{Single VIC Assumption:} Our annotation focused on identifying the primary vulnerability-inducing commit for each CVE. Some vulnerabilities may result from multiple contributing commits across different timeframes, which our current annotation does not capture exhaustively.

\section*{Ethical Considerations}

This work presents a benchmark dataset for evaluating vulnerability-inducing commit (VIC) detection tools. All CVEs and source code used in this benchmark are publicly available through the National Vulnerability Database~\citep{nvd2026} and open-source repositories. No private or proprietary data was collected.

We acknowledge the dual-use nature of security benchmarks: while VICBench is intended to advance the evaluation of defensive security tools, detailed vulnerability information could theoretically inform offensive use. To mitigate this risk, all CVEs included in this dataset are already publicly disclosed and patched. The benchmark does not introduce new vulnerability information beyond what is already publicly available.

No human subjects were involved in dataset construction. The annotation process was conducted by the authors and a security expert reviewer.

\bibliography{custom}

\appendix

\clearpage
\FloatBarrier

\section{Appendix}

\subsection{Additional Case Study: CVE-2019-15477}
\label{sec:appendix_case_1}

\begin{figure}[t]
\centering
\small
\begin{tikzpicture}[scale=0.55, transform shape,
    node distance=0.5cm,
    every node/.style={font=\small},
    commit/.style={rectangle, draw, minimum width=2.2cm, minimum height=0.8cm, align=center, fill=white},
    vic/.style={commit, fill=red!20, draw=red!60!black, thick},
    refactor/.style={commit, fill=blue!10, draw=blue!50!black},
    algo/.style={commit, fill=yellow!20, draw=orange!70!black, thick},
    fix/.style={commit, fill=green!20, draw=green!60!black, thick},
    timeline/.style={->, thick, gray!60},
    label/.style={font=\small, text width=2.2cm, align=center}
]

\draw[timeline, line width=1pt] (0,0) -- (12.8,0);

\foreach \x/\year in {0/2014, 3.0/2015, 6.0/2016, 9.0/2017, 12.8/2019} {
    \draw[gray!60] (\x,0.12) -- (\x,-0.12);
    \node[below, font=\small] at (\x,-0.3) {\year};
}

\node[vic] (vic) at (0.5, 2.3) {\textbf{d548655}\\Jun 2014};
\node[label, below=0.04cm of vic, text=red!70!black] {\textbf{Initial VIC}\\RouteHandler};

\node[refactor] (del) at (3.2, 2.3) {\textbf{2071f2e}\\Feb 2015};
\node[label, below=0.04cm of del, text=blue!70!black] {File Deletion\\to Err.java};

\node[algo] (llm) at (6, 3.6) {\textbf{49b524c}\\May 2015};
\node[label, below=0.04cm of llm, text=orange!80!black] {\textbf{LLM4SZZ}\\stopped};

\node[algo] (vszz) at (9.0, 3.6) {\textbf{8a3308c}\\Sep 2017};
\node[label, below=0.04cm of vszz, text=orange!80!black] {\textbf{V-SZZ}\\stopped};

\node[fix] (fix) at (12, 2.3) {\textbf{27b4af2}\\Aug 2019};
\node[label, below=0.04cm of fix, text=green!70!black] {\textbf{XSS Fix}\\Merge PR};

\draw[->, thick, red!60!black] (vic) -- (del);
\draw[->, thick, blue!60!black] (del) -- (llm);
\draw[->, thick, orange!70!black, dashed] (llm) -- (vszz);
\draw[->, thick, orange!70!black, dashed] (vszz) -- (fix);

\draw[very thick, red!50, dashed, line width=1.5pt] (0.5,0.3) -- (12.8,0.3);
\node[fill=white, font=\small, text=red!70!black] at (6.65, 0.3) {Vulnerability Persists};

\end{tikzpicture}
\caption{CVE-2019-15477 evolution timeline showing vulnerability persistence through file deletion and refactorings from Jun 2014 to Aug 2019.}
\label{fig:cve_timeline_15477}
\end{figure}
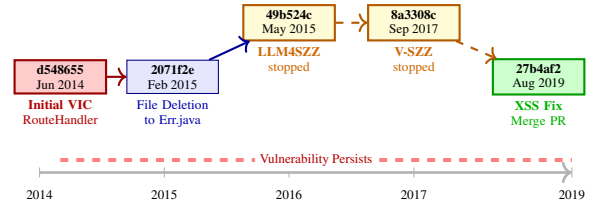

CVE-2019-15477 is an XSS vulnerability in Jooby's default error handler that allowed unescaped error messages to be rendered in HTML responses. This case demonstrates why automated VIC detection fails when vulnerabilities persist through file deletions and architectural refactorings. Figure~\ref{fig:cve_timeline_15477} shows the 5-year evolution of this vulnerability across major code transformations.

\paragraph{Commit Evolution History}

The vulnerability followed this evolution through the codebase:

\begin{enumerate}
    \item \textbf{Initial Introduction (d548655, Jun 2014):} Created \texttt{RouteHandler.java} with unescaped error model:
    \begin{lstlisting}[style=codestyle, basicstyle=\ttfamily\scriptsize]
private Map<String,Object> errorModel(
    Request request, Exception ex, HttpStatus status) {
    Map<String,Object> error = new LinkedHashMap<>();
    String message = ex.getMessage();
    message = message == null ? status.reason() : message;
    error.put("message", message);       // UNESCAPED
    error.put("stackTrace", dump(ex));
    error.put("status", status.value());
    error.put("reason", status.reason());  // UNESCAPED
    error.put("referer", request.header("Referer")...);
    return error;
}
    \end{lstlisting}

    \item \textbf{File Deletion (2071f2e, Feb 2015):} \texttt{RouteHandler.java} deleted; error logic moved to \texttt{Err.java} interface default method. The unescaped message handling persisted.

    \item \textbf{LLM4SZZ Stopped (49b524c, May 2015):} Refactored to use \texttt{toMap()} method in \texttt{DefHandler} class. The vulnerability remained:
    \begin{lstlisting}[style=codestyle, basicstyle=\ttfamily\scriptsize]
Results.when(MediaType.html, () ->
    Results.html(VIEW).put("err", ex.toMap()))
    \end{lstlisting}

    \item \textbf{V-SZZ Stopped (8a3308c, Sep 2017):} Added stacktrace parameter \texttt{toMap(stackstrace)}. Still unescaped:
    \begin{lstlisting}[style=codestyle, basicstyle=\ttfamily\scriptsize]
err.put("message", message);  // Still unescaped!
err.put("reason", status.reason());  // Still unescaped!
    \end{lstlisting}

    \item \textbf{Fix Applied (27b4af2, Aug 2019):} Merge commit integrating PR \#1368 that added XSS filtering after 5 years:
    \begin{lstlisting}[style=codestyle, basicstyle=\ttfamily\scriptsize]
Function<Object,String> xssFilter = env.xss("html")...;
details.compute("message", escaper);  // FIXED
details.compute("reason", escaper);   // FIXED
    \end{lstlisting}
\end{enumerate}

\paragraph{Algorithm Performance Analysis}

The \textbf{V-SZZ} stopped at commit 8a3308c (Sep 2017), which only added a boolean parameter for stacktrace visibility. Git blame cannot trace beyond the Feb 2015 file deletion, treating \texttt{Err.java} as unrelated to the deleted \texttt{RouteHandler.java}. \textbf{LLM4SZZ} stopped at commit 49b524c (May 2015), misidentifying architectural refactoring as vulnerability introduction and hallucinating a false root cause about "Supplier patterns." \textbf{VIC-Agent} identified d548655 (Jun 2014) through iterative backward search using \texttt{git log -S} to trace the vulnerability pattern (unescaped error messages) through 6 commits: 8a3308c (parameter addition), 49b524c (method refactoring), 2071f2e (package move), 3f19db103 (file extraction), and finally d548655 (origin). The approach tracked semantic continuity across 3 method names (\texttt{errorModel() → err() → toMap()}) and multiple files despite transformations that broke line-based and blame-based tracing.

\subsection{CWE Type Distribution}
\label{sec:appendix_cwe_dist}

Table~\ref{tab:cwe_distribution} shows the distribution of CWE types in our dataset.

\begin{table}[h]
\centering
\small
\begin{tabular}{lcl}
\toprule
\textbf{CWE} & \textbf{Count} & \textbf{Description} \\
\midrule
CWE-79 & 12 & Cross-site Scripting (XSS) \\
CWE-22 & 10 & Path Traversal \\
CWE-20 & 8 & Improper Input Validation \\
CWE-611 & 4 & XML External Entity (XXE) \\
CWE-601 & 4 & URL Redirection to Untrusted Site \\
\midrule
\textbf{Other (43)} & \textbf{63} & \textbf{Diverse types} \\
\bottomrule
\end{tabular}
\caption{Top 5 CWE types in our dataset. Total of 48 unique CWE types across 100 instances.}
\label{tab:cwe_distribution}
\end{table}

\subsection{Cohen's Kappa Calculation Details}
\label{sec:appendix_kappa}

Cohen's kappa is defined as:
\begin{equation}
\kappa = \frac{P_o - P_e}{1 - P_e}
\end{equation}
where $P_o$ is the observed agreement proportion and $P_e$ is the expected agreement by chance.

\paragraph{Agreement Criteria:} Two annotations were considered in agreement if and only if both identified the exact same commit. This strict definition avoids circular reasoning about ground truth.

For VIC annotation, the probability of random agreement depends on the commit pool size from which annotators could select. For each CVE, we identified the earliest VIC between the two annotations and counted commits from this earliest VIC (inclusive) to the fix commit (exclusive) using git history. Pool sizes ranged from 2 to 212,671 commits with a median of 1,690 commits. Expected agreement was computed as:
\begin{equation}
P_e = \frac{1}{N} \sum_{i=1}^{N} \frac{1}{n_i}
\end{equation}
where $N=100$ is the number of CVEs and $n_i$ is the commit pool size for CVE $i$.

In our case, $P_o = 0.71$ (71 agreements out of 100 cases) and $P_e = 0.0115$ (1.15\%), yielding $\kappa = 0.707$.

\end{document}